\documentclass[twocolumn,floatfix,showpacs,showkeys,preprintnumbers,nofootinbib,superscriptaddress]{revtex4}

\usepackage{graphicx}
\usepackage{amsmath}
\usepackage{amsfonts}
\usepackage{amssymb}
\usepackage{color}
\usepackage{subfigure}
\usepackage{epsfig}
\usepackage{morefloats}
\usepackage{multirow}
\usepackage{graphicx,booktabs}
\usepackage{mathrsfs}
\usepackage{txfonts}
\usepackage{indentfirst}
\usepackage{graphicx,booktabs}

\usepackage{longtable,lscape}
\usepackage[colorlinks, citecolor=blue,anchorcolor=red,menucolor=red, linkcolor=red,filecolor=red,urlcolor=blue,frenchlinks=red]{hyperref}

\newcommand{\vsig}{\mbox{\boldmath$\sigma$\unboldmath}}

\usepackage{comment}

\begin{document}

%\begin{spacing}{2.0}

\title{Toward discovering the excited $\Omega$ baryons through nonleptonic weak decays of $\Omega_c$}

\author{Kai-Lei Wang}\email{wangkaileicz@foxmail.com}

\affiliation{Department of Physics,
Changzhi University, Changzhi, Shanxi,046011,China}
\affiliation{Synergetic Innovation Center for Quantum Effects and
Applications (SICQEA), Hunan Normal University, Changsha 410081,
China}

\author{Qi-Fang L\"{u}}\email{lvqifang@hunnu.edu.cn}
\affiliation{Synergetic Innovation Center for Quantum Effects and
Applications (SICQEA), Hunan Normal University, Changsha 410081,
China}
\affiliation{Department of Physics, Hunan Normal University, and Key
Laboratory of Low-Dimensional Quantum Structures and Quantum Control
of Ministry of Education, Changsha 410081, China }

\author{Ju-Jun Xie}\email{xiejujun@impcas.ac.cn}
\affiliation{Institute of Modern Physics, Chinese Academy of
Sciences, Lanzhou 730000, China} \affiliation{School of Nuclear
Science and Technology, University of Chinese Academy of Sciences,
Beijing 101408, China} \affiliation{School of Physics and
Microelectronics, Zhengzhou University, Zhengzhou, Henan 450001,
China}

\author{Xian-Hui Zhong}\email{zhongxh@hunnu.edu.cn}
\affiliation{Synergetic Innovation Center for Quantum Effects and
Applications (SICQEA), Hunan Normal University, Changsha 410081,
China}
\affiliation{Department of Physics, Hunan Normal University, and Key
Laboratory of Low-Dimensional Quantum Structures and Quantum Control
of Ministry of Education, Changsha 410081, China }

\date{\today}

\begin{abstract}

The nonleptonic weak decay processes $\Omega_c \to
\Omega\pi^+/\Omega(1P)\pi^+/\Omega(1D)\pi^+/\Omega(2S)\pi^+$ are studied using the constituent quark model. The branching fraction of
$\Omega_c \to \Omega\pi^+$ is predicted to be $1.0\%$. Considering the newly observed $\Omega(2012)$ resonance as a conventional
$1P$-wave $\Omega$ excite state with spin-parity $J^P=3/2^-$, the newly measured ratio $\mathcal{B}[\Omega_c\to \Omega(2012)\pi^+ \to
(\Xi\bar{K})^-\pi^+ ]/\mathcal{B}[\Omega_c\to \Omega \pi^+]$ at Belle can be well
understood. Besides, the production rates for the missing $1P-$wave
state $\Omega (1^2P_{1/2^-})$, two spin quartet $1D-$wave states
$\Omega (1^4D_{1/2^+})$ and $\Omega (1^4D_{3/2^+})$, and two
$2S$-wave states $\Omega(2^2S_{1/2^+})$ and $\Omega(2^4S_{3/2^+})$
are also investigated. It is expected that these missing excited
$\Omega$ baryons should have large potentials to be discovered
through the nonleptonic weak decays of $\Omega_c$ in forthcoming
experiments by Belle II and/or LHCb.

\end{abstract}

\pacs{}

\maketitle

\section{Introduction}

Establishing a relatively complete hadron spectrum and understanding
the properties of hadrons are important topics in hadron physics.
The knowledge about the $\Omega$ baryon spectrum is very scarce. So
far, the ground state $\Omega(1672)$ and its four possible excited
states $\Omega(2012)$, $\Omega(2250)$, $\Omega(2380)$, and
$\Omega(2470)$, have been observed in
experiments~\cite{ParticleDataGroup:2020ssz}. The unambiguous
discovery of $\Omega(1672)$ in both production and decay was by
Barnes \emph{et al.} in 1964  using the $K^-$-meson beam at the
Brookhaven National Laboratory~\cite{Abrams:1964tu,Barnes:1964pd}.
In 1985, the $\Omega(2250)$ and $\Omega(2380)$ resonances decaying
into $\Xi^-\pi^+K^-$ were observed in an experiment at the CERN
SPS charged hyperon beam using incident
$\Xi^-$~\cite{Biagi:1985rn}. In 1987, the $\Omega(2250)$ resonance
was  produced in $K^-p$ interactions at SLAC~\cite{Aston:1987bb}. In
1988, the $\Omega(2470)$ resonance was observed in the
$\Omega^-\pi^+\pi^-$ invariant mass spectrum with a signal significance
claimed to be at least 5.5 standard deviations by using the $K^-p$
scattering at SLAC~\cite{Aston:1988yn}. Since then, there was no
progress toward searching for $\Omega$ resonances for as long as 30
years due to no effective production mechanisms. In order to promote
the experiment, people proposed to produce $\Omega$ states on a
proton target in CLAS12 through the photoproduction
processes~\cite{Afanasev:2012fh}, or produce them by using a
secondary kaon beam from the photoproduction processes at JLab
etc.~\cite{Amaryan:2015swp,Briscoe:2015qia}.

In 2018, the first low-lying $\Omega(2012)$ resonance was observed
by the Belle Collaboration in the $K^-\Xi^0$ and $K_S^0\Xi^-$
invariant mass distributions by using a data sample of $e^+e^-$
annihilations~\cite{Belle:2018mqs}. The $\Omega(2012)$ resonance may
favor the low-lying $P$-wave excited $\Omega$ state with
$J^P=3/2^-$~\cite{Xiao:2018pwe,Liu:2019wdr,Aliev:2018yjo,Aliev:2018syi,Polyakov:2018mow},
although it may be a candidate of  hadronic molecule state as
discussed in the literatures
~\cite{Wang:2007bf,Wang:2008zzz,Valderrama:2018bmv,Lin:2018nqd,Huang:2018wth,Pavao:2018xub,Lu:2020ste,Ikeno:2020vqv}.
Recently, the Belle Collaboration also discovered the $\Omega(2012)$
resonance by using the $\Omega_c$ weak decay process $\Omega_c\to
\Omega(2012)\pi^+$~\cite{Belle:2021gtf}. The measured branching
fraction ratio $\mathcal{B}[\Omega_c\to \Omega(2012)\pi^+ \to
(\Xi\bar{K})^-\pi^+ ]/\mathcal{B}[\Omega_c\to \Omega \pi^+]$ is
$0.220\pm0.059(\mathrm{stat.})\pm0.035(\mathrm{syst.})$~\cite{Belle:2021gtf}.
Such a large relative ratio indicates that the weak decay processes
$\Omega_c\to \Omega^*(X)\pi^+$ may provide a new and ideal platform
to investigate the low-lying excited states $\Omega^*(X)$ both
theoretically and experimentally.~\footnote{Here and after, we
donote $\Omega$ excited state as $\Omega^*(X)$ with mass $X$ in the
unit of MeV.}

\begin{table*}[htbp]
\begin{center}
\caption{\label{sp1} The predicted mass spectrum (MeV) of $\Omega$
baryons with principal quantum number $N \leq 2$  in various quark
models. The baryon states denoted as $|N_6, ^{2S+1}N_3, N, L,
J^P\rangle$, where $N_6$ stands for the irreducible representation
of spin-flavor SU(6) group, $N_3$ stands for the irreducible
representation of flavor SU(3) group, and $N$, $S$, $L$, and $J^P$
stand for the principal, spin, total orbital angular momentum, and
spin-parity quantum numbers, respectively. In the $L-S$ coupling
scheme, the $\Omega$ states are also denoted by $n^{2S+1}L_{J^P}$. }
\begin{tabular}{lccccccccccccccccccccccccccccccccccccccccccccc}\hline\hline
%State                            ~~~~ &             ~~~~ &Predicted                ~~~~&Predicted  ~~~~&Predicted   ~~~~&Predicted  ~~~~&Predicted  ~~~~&Predicted ~~~~&        \\
$n^{2S+1}L_{J^p}$  ~~&$|N_6,^{2S+1}N_3,N,L,J^P\rangle$
~~&Ref.~\cite{Oh:2007cr}  ~~&Ref.~\cite{Capstick:1985xss}
~~&Ref.~\cite{Faustov:2015eba}  ~~&Ref.~\cite{Chao:1980em}
~~&Ref.~\cite{Chen:2009de}  ~~&Ref.~\cite{Pervin:2007wa}
~~&Ref.~\cite{Engel:2013ig} ~~&Ref.~\cite{Liu:2019wdr}~~& Observed
mass \\ \hline
$1^4S_{\frac{3}{2}^+}$   ~~&$|56,^{4}10,0,0,\frac{3}{2}^+\rangle$~~&1694~~&1635~~&1678~~&1675~~&1673~~&1656~~&1642(17)~~&1672~~ &1672.45\\
$1^2P_{\frac{1}{2}^-}$   ~~&$|70,^{2}10,1,1,\frac{1}{2}^-\rangle$~~&1837~~&1950~~&1941~~&2020~~&2015~~&1923~~&1944(56)~~&1957~~ & \\
$1^2P_{\frac{3}{2}^-}$   ~~&$|70,^{2}10,1,1,\frac{3}{2}^-\rangle$~~&1978~~&2000~~&2038~~&2020~~&2015~~&1953~~&2049(32)~~&2012~~ &2012.5\\
$2^2S_{\frac{1}{2}^+}$   ~~&$|70,^{2}10,2,0,\frac{1}{2}^+\rangle$~~&2140~~&2220~~&2301~~&2190~~&2182~~&2191~~&2350(63)~~&2232~~ &\\
$2^4S_{\frac{3}{2}^+}$   ~~&$|56,^{4}10,2,0,\frac{3}{2}^+\rangle$~~&    ~~&2165~~&2173~~&2065~~&2078~~&2170~~&        ~~&2159~~ &\\
$1^2D_{\frac{3}{2}^+}$   ~~&$|70,^{2}10,2,2,\frac{3}{2}^+\rangle$~~&2282~~&2345~~&2304~~&2265~~&2263~~&2194~~&2470(49)~~&2245~~ & \\
$1^2D_{\frac{5}{2}^+}$   ~~&$|70,^{2}10,2,2,\frac{5}{2}^+\rangle$~~&    ~~&2345~~&2401~~&2265~~&2260~~&2210~~&        ~~&2303~~ &   \\
$1^4D_{\frac{1}{2}^+}$   ~~&$|56,^{4}10,2,2,\frac{1}{2}^+\rangle$~~&2140~~&2255~~&2301~~&2210~~&2202~~&2175~~&2481(51)~~&2141~~ & \\
$1^4D_{\frac{3}{2}^+}$   ~~&$|56,^{4}10,2,2,\frac{3}{2}^+\rangle$~~&2282~~&2280~~&2304~~&2215~~&2208~~&2182~~&2470(49)~~&2188~~ &   \\
$1^4D_{\frac{5}{2}^+}$   ~~&$|56,^{4}10,2,2,\frac{5}{2}^+\rangle$~~&    ~~&2280~~&2401~~&2225~~&2224~~&2178~~&        ~~&2252~~ & \\
$1^4D_{\frac{7}{2}^+}$   ~~&$|56,^{4}10,2,2,\frac{7}{2}^+\rangle$~~&    ~~&2295~~&2332~~&2210~~&2205~~&2183~~&        ~~&2321~~ &   \\
\hline\hline
\end{tabular}
\end{center}
\end{table*}

Theoretical studies on the $\Omega^*(X)$
resonances mainly focus on the mass spectrum within various
approaches, such as nonrelativistic quark
models~\cite{Kalman:1982ut,Menapara:2021vug,Pervin:2007wa,Liu:2019wdr,Chao:1980em,Chen:2009de},
relativistic quark
models~\cite{Capstick:1985xss,Faustov:2015eba,Loring:2001kx,Santopinto:2014opa},
Lattice QCD~\cite{Engel:2013ig,CLQCD:2015bgi}, and the Skyrme
model~\cite{Oh:2007cr}. The predicted mass spectrum for the
conventional $\Omega$ baryons are collected in Table~\ref{sp1} as a
reference. It can be seen that most of the predicted masses for the
$1P$-, $2S$- and $1D$-wave states lies in the mass ranges $\sim 2000
\pm 50$, $\sim 2200 \pm 50$, and $\sim 2300 \pm 50$ MeV,
respectively. Additionally, in
Refs.~\cite{An:2013zoa,An:2014lga,Yuan:2012zs}, the authors
investigated the low-lying five-quark $\Omega$ configurations with
negative parity and further considered their mixing combined the
corresponding low-lying three-quark $\Omega$ configurations.
Recently, stimulated by the newly observed resonance $\Omega(2012)$
at Belle, the strong decay behaviors of some low-lying $1P$-, $2S$-
and $1D$-wave $\Omega$ resonances were also systematically
investigated using the chiral quark
model~\cite{Xiao:2018pwe,Liu:2019wdr} and $^3P_0$
model~\cite{Wang:2018hmi}. The results suggest that the $1P$-, $2S$-
and $1D$-wave $\Omega$ baryons have  relatively narrow decay widths
of less than 50 MeV, and they may be discovered in the $\Xi \bar{K}$
and/or $\Xi(1530) \bar{K}$ final states. Some previous studies of the
decays can be found in the
Refs.~\cite{Bijker:2000gq,Bijker:2015gyk}.

On the other hand, there are only a few studies on the productions of
$\Omega$ and its excited states through the weak decays of
$\Omega_c$ in theory. For example, the productions of the ground state
$\Omega(1672)$ have been studied via semileptonic decays of
$\Omega_c$ using a constituent quark model~\cite{Pervin:2006ie} and
the nonleptonic two-body decays of $\Omega_c$ by the covariant
confined quark model~\cite{Gutsche:2018utw,Korner:1992wi} and the
light-front quark model~\cite{Hsiao:2020gtc}. In
Ref.~\cite{Pervin:2006ie}, the author also studied the productions
of the $1P$-wave excited states $\Omega^*(1P)$, which are considered
via the $\Omega_c$ semileptonic weak decay processes using a quark
model. On the other hand, the newly observed $\Omega(2012)$
resonance as a dynamically generated state was theoretically studied
in the nonleptonic weak decays of $\Omega_c\to \pi^+\Omega(2012)\to
(\Xi\bar{K})^-\pi^+ $ and $(\Xi\bar{K}\pi)^-\pi^+$ in
Ref.~\cite{Zeng:2020och}. So far, the productions of the $1P$-,
$2S$- and $1D$-wave excited states $\Omega(X)$ via the $\Omega_c$
nonleptonic weak decay processes are not systematically studied in
theory.

In this work, we systematically study the production of the
low-lying $1P$-, $2S$- and $1D$-wave resonances $\Omega^*(X)$ via
the hadronic weak decays of $\Omega_c\to \Omega^{(*)}(X)\pi^+$ using
the constituent quark model. Recently, this model has been
developed to study the hadronic weak decays of $\Lambda_c$, the
heavy quark conserving weak decays of $\Xi_Q$, and hyperon weak
radiative decay by Niu \emph{et
al.}~\cite{Niu:2020gjw,Niu:2021qcc,Niu:2020aoz}. This model is
similar to that developed to deal with the semileptonic decays of
heavy $\Lambda_Q$ and $\Omega_Q$ baryons in
Refs.~\cite{Pervin:2006ie,Pervin:2005ve}.

This paper is organized as follows. We perform the detailed
formalism of two-body nonleptonic weak decays of $\Omega_c$ in
Sec.~\ref{MODEL}. Then, the theoretical numerical results and
discussions are presented in Sec.~\ref{DISSCUS}. Finally, a short
summary is given in Sec.~\ref{SUM}.

\section{framework}\label{MODEL}

\subsection{The model}

\begin{figure}[htbp]
\centering
\includegraphics[width=0.4\textwidth]{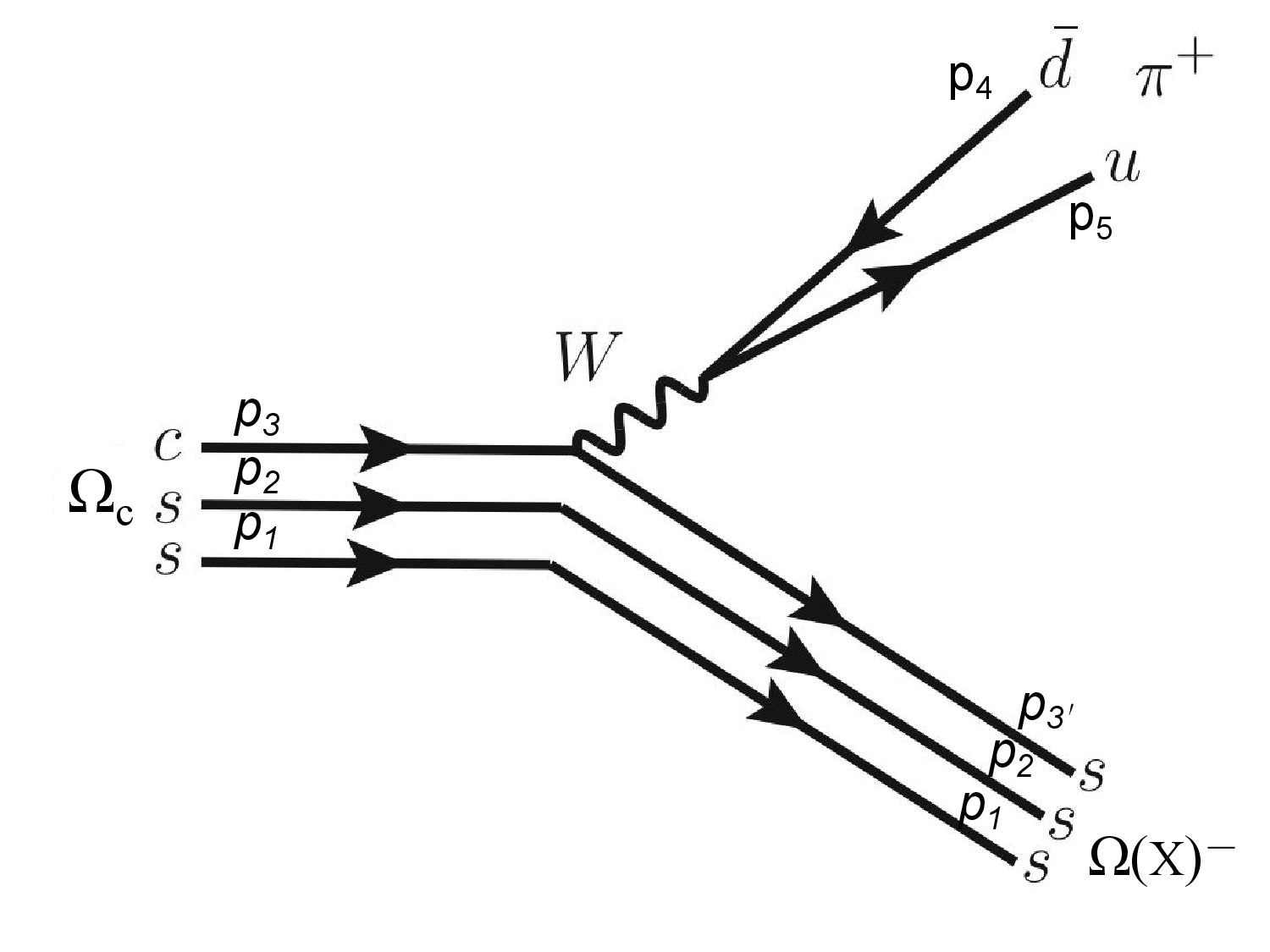}
\caption{The nonleptonic weak decay Feynman diagram for the processes of $\Omega_c\rightarrow \Omega(X)^- \pi^+$.}\label{tu}
\end{figure}

A unique feature of $\Omega_c \to \Omega^{(*)}(X)\pi^+$ is that this
decay proceeds only via external $W$-emission
diagram~\cite{Cheng:2021qpd}, which is displayed in Fig.~\ref{tu}.
We consider the simple quark-level transition $c\to s u \bar{d}$,
which is relevant for the Cabibbo-favored decay process of
$\Omega_c\to\Omega(X)^-\pi^+$. The effective Hamiltonian for $c\to s
u \bar{d}$ can be given by~\cite{Buchalla:1995vs}
\begin{eqnarray}\label{dww}
H_W&=&\frac{G_F}{\sqrt{2}}V_{cs}^* V_{ud} (C_1\mathcal{O}_1+C_2\mathcal{O}_2),
\end{eqnarray}
where $G_F=1.1663787\times 10^{-5}$ GeV$^{-2}$ is the Fermi constant~\cite{ParticleDataGroup:2020ssz}, and $C_1=1.26$ and $C_2=-0.51$
are the Wilson coefficients taken at the $m_c$ scale~\cite{Buchalla:1995vs}. The
Cabibbo-Kobayashi-Maskawa matrix elements $V_{cs}=0.987 $ and
$V_{ud}= 0.974$ are taken from the Review of Particle Physics
(RPP)~\cite{ParticleDataGroup:2020ssz}, and the current-current
operators are
\begin{eqnarray}\label{dww}
\mathcal{O}_1&=&\bar{\psi}_{\bar{s}_a}\gamma_\mu(1-\gamma_5)\psi_{c_a}\bar{\psi}_{\bar{u}_b}\gamma^\mu(1-\gamma_5)\psi_{d_b},\\
\mathcal{O}_2&=&\bar{\psi}_{\bar{s}_a}\gamma_\mu(1-\gamma_5)\psi_{c_b}\bar{\psi}_{\bar{u}_b}\gamma^\mu(1-\gamma_5)\psi_{d_a},
\end{eqnarray}
with $\psi_{j_\delta}$ ($j=u/d/s/c$, $\delta=a/b$) representing the $j$th quark field in a meson or baryon, and $a$ and $b$ being color indices.

According to its parity behavior, $H_W$ can be separated into a parity-conserving part ($H_{W}^{PC}$) and a parity-violating part ($H_{W}^{PV}$)~\cite{Niu:2020gjw},
\begin{eqnarray}\label{Hww}
H_{W}=H_{W}^{PC}+H_{W}^{PV}.
\end{eqnarray}
With a non-relativistic expansion, the two operators can be approximately expressed as~\cite{Niu:2020gjw}
\begin{eqnarray}\label{Hw}
H_{W}^{PC}&\simeq&\frac{G_F}{\sqrt{2}}V_{cs}^* V_{ud} \frac{C_i \phi_c^i \gamma}{(2\pi)^3}\delta^3(\textbf{p}_3-\textbf{p}_3'-\textbf{p}_4-\textbf{p}_5)\{\langle s_3'|I|s_3\rangle \nonumber \\
&&\langle s_5 \bar{s}_4|\vsig|0\rangle\left(\frac{\textbf{p}_5}{2m_5}+\frac{\textbf{p}_4}{2m_4}\right)-\Big [\left(\frac{\textbf{p}_3'}{2m_3'}+\frac{\textbf{p}_3}{2m_3}\right)\langle s_3'|I|s_3\rangle  \nonumber \\
&&-i\langle s_3'|\vsig|s_3\rangle\times \left(\frac{\textbf{p}_3}{2m_3}-\frac{\textbf{p}_3'}{2m_3'}\right)\Big ]\langle s_5 \bar{s}_4|\vsig|0\rangle   \nonumber \\
&&-\langle s_3'|I|s_3\rangle \Big[\left(\frac{\textbf{p}_5}{2m_5}+\frac{\textbf{p}_4}{2m_4}\right)\langle s_5 \bar{s}_4|\vsig|0\rangle-i\langle s_5 \bar{s}_4|\vsig|0\rangle  \nonumber \\
&&\times \left(\frac{\textbf{p}_4}{2m_4}-\frac{\textbf{p}_5}{2m_5}\right)\Big]+\langle s_3'|I|s_3\rangle\left(\frac{\textbf{p}_3'}{2m_3'}+\frac{\textbf{p}_3}{2m_3}\right) \nonumber \\
&&\langle s_5 \bar{s}_4|I|0\rangle \}\hat{\alpha}_3^{-},\\
H_{W}^{PV}&\simeq&\frac{G_F}{\sqrt{2}}V_{cs}^* V_{ud} \frac{C_i \phi_c^i \gamma}{(2\pi)^3}\delta^3(\textbf{p}_3-\textbf{p}_3'-\textbf{p}_4-\textbf{p}_5)\{-\langle s_3'|I|s_3\rangle \nonumber \\
&&\langle s_5 \bar{s}_4|I|0\rangle- \langle s_3'|\vsig|s_3\rangle\langle s_5 \bar{s}_4|\vsig|0\rangle)\hat{\alpha}_3^{-}.
\end{eqnarray}
In the above equations, $\textbf{p}_j$ and $m_j$ stand for the
momentum and mass of the $j$th quark, respectively, as shown in
Fig.~\ref{tu}. The $\phi_c^i$ ($i=1,2$ and $\phi_c^1=1$,
$\phi_c^2=\frac{1}{3}$) are color factors, $I$ is the dimension-two unit
matrix, and $\hat{\alpha}_3^{-}$ is the flavor operator which
transforms $c$ quark to $s$ quark. The $s_j$ and $\bar{s}_4$ stand
for the spin of the $j$th quark and the fourth antiquark,
respectively. $\gamma$ is a symmetry factor and equals to one for a
direct pion emission process considering in present work.

In order to evaluate the spin matrix element $\langle s_5
\bar{s}_4|I|0\rangle$ and $\langle s_5 \bar{s}_4|\vsig|0\rangle$
including an antiquark,  the particle-hole
conjugation~\cite{Racah:1942gsc} should be employed. Within the
particle-hole conjugation relation:
\begin{eqnarray}
|j,-m\rangle\rightarrow(-1)^{j+m}|j,m\rangle \, ,
\end{eqnarray}
the antiquark spin transforms as follows: $|\bar{\uparrow}\rangle
\rightarrow |\downarrow\rangle$ and $|\bar{\downarrow}\rangle
\rightarrow -|\uparrow\rangle$. For instance:
\begin{eqnarray}
&& \langle
\frac{1}{\sqrt{2}}(\uparrow_5\bar{\downarrow}_4-\downarrow_5\bar{\uparrow}_4)|I|0\rangle
= \frac{1}{\sqrt{2}}\left(
\langle\uparrow_5|I|-\uparrow_4\rangle-\langle\downarrow_5|I|\downarrow_4\rangle
\right) \nonumber \\
&& =-\sqrt{2}.
\end{eqnarray}

For a given decay process $A\to BC$, the transition amplitude $\mathcal{M}$ is calculated by
\begin{eqnarray}\label{Hww}
\mathcal{M}_{J_f,J_f^z;J_i,J_i^z}&=&\langle C(\textbf{P}_f;J_f,J_f^z)B(\textbf{q})|H_{W}|A(\textbf{P}_i;J_i,J_i^z)\rangle, \nonumber  \\
&=&\langle C(\textbf{P}_f;J_f,J_f^z)B(\textbf{q})|H_{W}^{PC}|A(\textbf{P}_i;J_i,J_i^z)\rangle  \nonumber  \\
&&+\langle C(\textbf{P}_f;J_f,J_f^z)B(\textbf{q})|H_{W}^{PV}|A(\textbf{P}_i;J_i,J_i^z)\rangle.  \nonumber  \\
&=&\mathcal{M}_{J_f^z,J_i^z}^{PC}+\mathcal{M}_{J_f^z,J_i^z}^{PV},
\end{eqnarray}
where $A(\textbf{P}_i;J_i,J_i^z)$, $B(\textbf{q})$ and
$C(\textbf{P}_f;J_f,J_f^z)$ stands for the wave functions of the
initial baryon $A$, final meson $B$ and final baryon $C$,
respectively. $(\mathbf{P}_i,\mathbf{P}_f)$, $(J_i,J_f)$, and
$(J_i^z,J_f^z)$  are the momentum, the total angular momentum and
the third component of the total angular momentum of the initial
baryon $A$ and the final baryon $C$, respectively. $\textbf{q}$ is the three-momentum of the final
state meson in the initial state rest frame.

Then, the partial decay width for a given decay process $A\to BC$ can be expressed as
\begin{equation}\label{dww}
\Gamma=\frac{\Phi(ABC)}{2J_A+1}\sum_{\texttt{spins}}|\mathcal{M}|^2,
\end{equation}
where $\Phi(ABC)$ is the phase-space factor for the decay.

The choice of phase space is not clear. For the phase space factor $\Phi(ABC)$, there are three typical options adopted in
the literature~\cite{Kokoski:1985is,Kumano:1988ga,Geiger:1994kr,Capstick:2000qj}. The usual option is the relativistic phase-space factor (RPF)
\begin{equation}\label{dww}
\Phi(ABC)=8\pi^2\frac{|\textbf{q}|E_BE_C}{M_A},
\end{equation}
where $M_A$ is the mass of the initial hadron $A$, while $E_B$ and
$E_C$ stand for the energies of final hadrons $B$ and $C$, respectively.

To match the transition matrix element calculated
non-relativistically, a fully nonrelativistic phase-space factor (NRPF) is
used, that is
\begin{equation}\label{dwb}
\Phi(ABC)=8\pi^2\frac{|\textbf{q}|M_BM_C}{M_A},
\end{equation}
where $M_B$ and $M_C$ is the mass of the final hadron $B$ and $C$, respectively.

However, in many cases the momenta of the final hadrons are quite large so that the relativistic phase space
is significantly different from the nonrelativistic limit. In Ref.~\cite{Kokoski:1985is}, Kokoski and Isgur
suggested a ``mock-hadron'' phase-space factor (MHPF),
\begin{equation}\label{dwb}
\Phi(ABC)=8\pi^2\frac{|\textbf{q}|\tilde{M}_B\tilde{M}_C}{\tilde{M}_A},
\end{equation}
in their calculation of meson decay widths.
The $\tilde{M}_A$, $\tilde{M}_B$ and $\tilde{M}_C$  are effective hadron masses of hadron $A$, $B$ and $C$, respectively.
They are evaluated with a spin-independent inter-quark interaction. In the weak-binding limit,
the mass of $\pi$ meson is degenerate with that of $\rho$ meson.

\subsection{Wave functions}

To work out the decay amplitude $\mathcal{M}$, we need the wave functions of
the initial and final states. Here, the initial state is the ground
$\Omega_c$ baryon, the final states are the $\pi^+$ meson and the
$\Omega^{(*)}(X)$ states. These wave functions are constructed
within the non-relativistic constituent quark model. For simplicity,
the spatial wave functions of the baryons and mesons are adopted the
harmonic oscillator form in our calculations.

The spatial wave function for a baryon with principal quantum number $N$, total orbital angular momentum quantum numbers $L$, and $M_L$
is a product of the $\rho$-oscillator part and the $\lambda$-oscillator part.
In momentum space, the baryon spatial wave function is given by~\cite{Liu:2019wdr}
\begin{eqnarray}\label{Hk}
\Psi^{\sigma}_{NLM_L}(\textbf{p}_\rho,\textbf{p}_\lambda) \! = \! \!
\! \sum_{N,M_L} C^{n_\rho l_\rho m_\rho}_{n_\lambda l_\lambda
m_\lambda}\left[\psi_{n_\rho l_\rho m_\rho}(\textbf{p}_\rho)
\psi_{n_\lambda l_\lambda
m_\lambda}(\textbf{p}_\lambda)\right]^{\sigma}_{NLM_L},
\end{eqnarray}
with $N=2(n_{\rho}+n_{\lambda})+l_{\rho}+l_{\lambda}$, $M_L =
m_\rho+m_\lambda$, and
\begin{eqnarray}\label{Hww}
\psi^{\alpha}_{n l m}(\textbf{p})=(i)^l(-1)^n\left[\frac{2n!}{(n+l+1/2)!}\right]^{1/2}\frac{1}{\alpha^{l+3/2}}  \nonumber \\
\mathrm{exp}\left(-\frac{\textbf{p}^2}{2\alpha^2}\right)L_n^{l+1/2}(\textbf{p}^2/\alpha^2)\mathcal{Y}_{lm}(\textbf{p}).
\end{eqnarray}
Here,
$\mathcal{Y}_{lm}(\textbf{p})=|\textbf{p}|^{l}Y_{lm}(\mathbf{\hat{p}})$
is $l$th solid harmonic polynomial. $\textbf{p}_\rho$ and
$\textbf{p}_\lambda$ are the internal momenta of the $\rho$- and
$\lambda$-oscillator wave functions, respectively. They can be
expressed as functions of the quark momenta $\mathbf{p}_j$
($j=1,2,3$):
\begin{eqnarray}
\mathbf{p}_\rho&=&\frac{\sqrt{2}}{2}(\textbf{p}_1-\textbf{p}_2),\\
\mathbf{p}_\lambda&=&\frac{\sqrt{6}}{2}\frac{m_3(\textbf{p}_1+\textbf{p}_2)-(m_1+m_2)\textbf{p}_3}{m_1+m_2+m_3}.
\end{eqnarray}
The $n_\rho$ and $n_\lambda$ are the principal quantum numbers of
the $\rho$- and $\lambda$-mode oscillators, respectively.
$(l_\rho,m_\rho)$ and $(l_\lambda,m_\lambda)$ are the orbital
angular momentum quantum numbers of the $\rho$- and $\lambda$-mode
oscillators, respectively. $\sigma = s, \rho, \lambda, a, ...$ stand
for different excitation modes with different permutation
symmetries. $\alpha_\rho$ and $\alpha_\lambda$ are two oscillator
parameters. For the $\Omega$ baryons, we have
$\alpha_\rho=\alpha_\lambda$, while for the charmed $\Omega_c$
baryons, we have
\begin{eqnarray}
\alpha_{\lambda}=\left(\frac{3m_c}{2m_s+m_c}\right)^{1/4}\alpha_{\rho},
\end{eqnarray}
where $m_s$ and $m_c$ stand for the masses of the strange and
charmed quarks, respectively. The flavor and spin wave functions of
the $\Omega_c$ and $\Omega$ baryons have been given in our previous
works~\cite{Xiao:2013xi,Wang:2017kfr}. The product of spin, flavor,
and spatial wave functions of the heavy baryons must be symmetric
since the color wave function is antisymmetric. The details about
the quark model classifications for the $\Omega_c$ spectrum can be
found in the works of~\cite{Wang:2017kfr,Yao:2018jmc,Zhong:2007gp},
while for the $\Omega$ baryon spectrum can be found in
Refs.~\cite{Xiao:2013xi,Liu:2019wdr}.

Finally, the wave function of the $\pi^+$ meson is constructed by
\begin{eqnarray}
\varphi(\textbf{p}_4,\textbf{p}_5)=\phi_{\pi^+}\chi^a\psi(\textbf{p}_4,\textbf{p}_5),
\end{eqnarray}
where the spin wave function $\chi^a$ is
\begin{eqnarray}
\chi^a=\frac{1}{\sqrt{2}}(\uparrow\downarrow-\downarrow\uparrow),
\end{eqnarray}
and the flavor wave function $\phi_{\pi^+}$ is
\begin{eqnarray}
\phi_{\pi^+}=u\bar{d}.
\end{eqnarray}
The spatial wave function in the momentum space is adopted the
simple harmonic oscillator form
\begin{eqnarray}\label{Hww}
\psi(\textbf{p}_4,\textbf{p}_5)=\frac{1}{\pi^{3/4}\beta^{3/2}}\mathrm{exp}\left[-\frac{(\textbf{p}_4-\textbf{p}_5)^2}{8\beta^2}\right],
\end{eqnarray}
where $\beta$ is a size parameter of the meson wave function. The $\textbf{p}_4$ and $\textbf{p}_5$
stand for the quark momenta of the $\pi^+$ meson as shown in Fig.~\ref{tu}.

\subsection{Parameters}

For self consistency, the quark model parameters are taken the same
as those adopted in our previous work~\cite{Wang:2017kfr}. The
constituent masses for the $u/d$, $s$ and $c$ quarks are taken to be
$m_{u/d}=330$ MeV, $m_{s}=450$ MeV and $m_{c}=1480$ MeV,
respectively. For the initial state $\Omega_c$, the harmonic
oscillator parameter $\alpha_{\rho}$ is taken to be
$\alpha_{\rho}=440$ MeV, the other harmonic oscillator parameter
$\alpha_{\lambda}$ is related to $\alpha_{\rho}$ by
$\alpha_{\lambda}=[3m_c/(2m_s+m_c)]^{1/4}\alpha_{\rho}$. For the
final state $\Omega^{(*)}(X)$, a unified harmonic oscillator
parameter is adopted, i.e., $\alpha_{\lambda}=\alpha_{\rho}=440$
MeV. For the $\pi^+$ meson, the size parameter is taken to be
$\beta=280$ MeV as that adopted in Ref.~\cite{Niu:2020gjw}. The
masses for the $\pi^+$, $\Omega$ and $\Omega_c$ are taken the RPP
average values 140 MeV, 1672 MeV and 2695 MeV,
respectively~\cite{ParticleDataGroup:2020ssz}. In the MHPF defined in Eq.~(\ref{dwb}),
we need determine the effective masses of the mock hadrons.
For the process $\Omega_c\to \Omega^{(*)}\pi$, we adopt $\tilde{M}_\pi=0.72$ GeV,
consistent with Kokoski and Isgur~\cite{Kokoski:1985is}, and $\tilde{M}_{\Omega_c} = M_{\Omega_c}$
and $\tilde{M}_{\Omega_c^{(*)}} = M_{\Omega_c^{(*)}}$.

\section{Numerical results and discussion}\label{DISSCUS}

In this work, considering the uncertainties from the relativistic effect,
we perform our calculations with the three typical phase space options,
RPF, NRPF and MHPF. Our results are listed in Table~\ref{result}.
It is seen that the nonleptonic weak decay properties of $\Omega_c$
have a significance dependence on the options of the phase space factor.
The results from RPF and MHPF are comparable with each other.
However, the predicted partial widths with NRPF are a factor of
$\sim2-6$ smaller those calculated with RPF and MHPF.

\begin{figure}[htbp]
\includegraphics[width=0.46\textwidth]{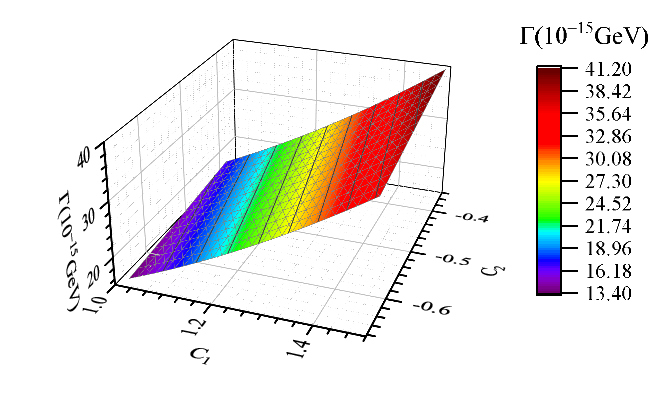}
\vspace{-0.5cm} \caption{The dependencies of the partial decay width of
$\Omega_c\rightarrow \Omega^- \pi^+$ on the parameters $C_1$ and $C_2$.
The results are obtained by adopting the RPF. }\label{picture}
\end{figure}

The Wilson coefficients $C_1$ and $C_2$ are usually taken to be $C_1=1.26$ and $C_2=-0.51$
at the $m_c$ scale~\cite{Buchalla:1995vs}.
These coefficients have some uncertainties due to their scale dependencies.
To see the effects of the uncertainties of $C_1$ and $C_2$ on our results,
as an example in Fig.~\ref{picture} we plot the partial width of $\Gamma[\Omega_c^0\to
\Omega^- \pi^+]$ as a function of the $C_1$ and $C_2$ in the range of $C_1\in(1.0,1.5)$
and $C_2\in(-0.64,-0.38)$.
From the figure, one can see that considering a $20\%$ uncertainty for
the Wilson coefficients $C_1=1.26$ and $C_2=-0.51$ at the $m_c$ scale,
the partial decay width of $\Gamma[\Omega_c^0\to
\Omega^- \pi^+]$ lies in the range of $(1.3 ,4.0)\times 10^{-14}$ GeV,
which shows a sizeable decadency on the Wilson coefficients.

\subsection{$\Omega_c^0\rightarrow \Omega^- \pi^+$}

First, we study the weak decay process $\Omega_c^0\rightarrow
\Omega^- \pi^+$. This weak decay process, as an
important process, has been widely studied by the Belle, BaBar,
CLEO, SELEX, FOCUS
collaborations~\cite{Belle:2017szm,BaBar:2007jdg,CLEO:2000dhf,FOCUS:2003ylx,Solovieva:2008fw,SELEX:2007bim,BaBar:2005yiy}.
With the RPF, the partial decay width of $\Omega_c^0\rightarrow
\Omega^- \pi^+$ is predicted to be
\begin{eqnarray}\label{aa}
\Gamma[\Omega_c^0\rightarrow \Omega^- \pi^+] &\simeq& 2.6 \times 10^{-14}~\text{GeV}.
\end{eqnarray}
By using the measured lifetime $\tau = 2.68 \times 10^{-13} s$ of
$\Omega_c^0$~\cite{ParticleDataGroup:2020ssz}, we further predict the branching fraction
\begin{eqnarray}\label{bb}
\mathcal{B}[\Omega_c^0\rightarrow \Omega^- \pi^+] \simeq 1.05\%.
\end{eqnarray}
If adopting the MHPF, there is a $\sim20\%$ correction to the
results of RMF. However, when adopting the NRPF the results are about a
factor of $\sim6.6$ smaller than that predicted with RMF.
From Table~\ref{s-wave}, it is found that our predicted branching fraction with both RPF and MHPF is close to the
predictions in Refs.~\cite{Cheng:1996cs,Hsiao:2020gtc}. While,
if adopting the NRPF, our predicted branching fraction $\mathcal{B}[\Omega_c^0\rightarrow \Omega^- \pi^+] \simeq 0.16\%$
is consistent with that from the covariant confined quark model~\cite{Gutsche:2018utw}.

\begin{table}[htbp]
\begin{center}
\caption{\label{s-wave} Predicted branching fraction for the
$\Omega_c^0\rightarrow \Omega^- \pi^+$ precess compared with that of
other theoretical works.}
\begin{tabular}{lccccccccccccccccccccccccccccccccccccccccccccc}\hline\hline
%State                            ~~~~ &             ~~~~ &Predicted                ~~~~&Predicted  ~~~~&Predicted   ~~~~&Predicted  ~~~~&Predicted  ~~~~&Predicted ~~~~&        \\
       &RPF/MHPF/NRPF   &Ref.~\cite{Gutsche:2018utw} &Ref.~\cite{Cheng:1996cs}  &Ref.~\cite{Hsiao:2020gtc}  &Ref.~\cite{Korner:1992wi} & \\ \hline
 ~~~~&1.05\%/0.82\%/0.16\% &0.2\%   &1.0\%  &0.5\%     &2.3\% &\\
\hline\hline
\end{tabular}
\end{center}
\end{table}

Furthermore, combined the predicted branching fraction
of $\mathcal{B}[\Omega_c^0\rightarrow \Omega^- \pi^+]$
with the measured relative branching ratios
$\frac{\Gamma[\Omega_c^0\rightarrow
\Xi^0\bar{K}^-\pi^+]}{\Gamma[\Omega_c^0\rightarrow \Omega^-
\pi^+]}=1.20\pm0.24$ and  $\frac{\Gamma[\Omega_c^0\rightarrow
\Xi^-\bar{K}^0\pi^+]}{\Gamma[\Omega_c^0\rightarrow \Omega^-
\pi^+]}=2.12\pm0.38$, the branching fractions for the three-body weak decay processes
$\Omega_c^0\rightarrow \Xi^0\bar{K}^-\pi^+/\Xi^-\bar{K}^0\pi^+$ can be obtained easily.
With the RPF, we have
\begin{eqnarray}
\mathcal{B}[\Omega_c^0\rightarrow \Xi^0\bar{K}^-\pi^+]&\simeq& (1.26 \pm 0.27) \times 10^{-2},\label{a1}\\
\mathcal{B}[\Omega_c^0\rightarrow \Xi^-\bar{K}^0\pi^+]&\simeq& (2.23
\pm 0.43) \times 10^{-2}\label{a2}.
\end{eqnarray}
While when adopting the NRPF,
we have small branching fractions
\begin{eqnarray}
\mathcal{B}[\Omega_c^0\rightarrow \Xi^0\bar{K}^-\pi^+]&\simeq& (0.19 \pm 0.04) \times 10^{-2},\label{a1}\\
\mathcal{B}[\Omega_c^0\rightarrow \Xi^-\bar{K}^0\pi^+]&\simeq& (0.33
\pm 0.06) \times 10^{-2}\label{a2},
\end{eqnarray}
due to the small nonrelativistic phase space factor.

\subsection{$\Omega_c^0\rightarrow \Omega^-(1P) \pi^+$ }\label{1p1}

In the $\Omega$ family, there are two $1P$-wave states $ \Omega
(1^2P_{1/2^-}) $ and $\Omega (1^2P_{3/2^-}) $ with spin-parity
$J^P=1/2^-$ and $J^P=3/2^-$, respectively. The newly observed
$\Omega(2012)$ resonance may favor the assignment of $\Omega
(1^2P_{3/2^-})$ state, since both the measured mass and width are
consistent with the quark model
predictions~\cite{Xiao:2018pwe,Liu:2019wdr,Aliev:2018yjo,Aliev:2018syi,Polyakov:2018mow}.
The masses of the unestablished $\Omega^{*}(X)$ states are taken the
predictions in Ref.~\cite{Liu:2019wdr}, which have been collected in
Table~\ref{sp1}. However, the $ \Omega (1^2P_{1/2^-}) $
classified in the quark model is still missing.

\begin{table*}[htbp]
\begin{center}
\caption{\label{result}  Predicted decay properties of the $\Omega_c
\rightarrow \Omega^{(*)}(X)^-\pi^+$ processes within three options of the phase space, RPF, NRPF and MHPF, respectively. $\Gamma_i$ stands for
the partial decay width, $\mathcal{B}$ stands for the branching
fraction, and $M_f$ stands for the mass of the final state
$\Omega^{(*)}(X)$.  The total width of $\Omega_c$ is $\Gamma =
2.47\times10^{-12}$ GeV (corresponding to life time $\tau = 2.68
\times10^{-13}\mathrm{s}$~\cite{ParticleDataGroup:2020ssz} ). The units for decay width $\Gamma_{i}$ and branching ratio $\mathcal{B}$ are $10^{-15}$ GeV and $10^{-3}$, respectively. }
\begin{tabular}{lccccccccccccccccccccccccccccccccccccccccccccc}\hline\hline
&~~~~~~~~~~~~~~&\multicolumn{3}{c}{\underline{~~~~~~~~~~~~~~~~~~~~~~RPF~~~~~~~~~~~~~~~~~~~~~~}}&\multicolumn{3}{c}{\underline{~~~~~~~~~~~~~~~~~~~~~~NRPF~~~~~~~~~~~~~~~~~~~~~~}}
&\multicolumn{3}{c}{\underline{~~~~~~~~~~~~~~~~~~~~~~MHPF~~~~~~~~~~~~~~~~~~~~~~}}\\
final state &$M_f$ (MeV)  &$\Gamma_i $   &$\mathcal{B}$   &$\frac{\Gamma[\Omega_c^0\to\Omega^{(*)}(X)^- \pi^+]}{\Gamma[\Omega_c^0\rightarrow \Omega^-
\pi^+]}$    &$\Gamma_i $   &$\mathcal{B}$    &$\frac{\Gamma[\Omega_c^0\to\Omega^{(*)}(X)^- \pi^+]}{\Gamma[\Omega_c^0\rightarrow \Omega^-
\pi^+]}$    &$\Gamma_i $   &$\mathcal{B}$    &$\frac{\Gamma[\Omega_c^0\to\Omega^{(*)}(X)^- \pi^+]}{\Gamma[\Omega_c^0\rightarrow \Omega^-
\pi^+]}$\\ \hline
$\Omega(1^4S_{\frac{3}{2}^+})\pi^+$ &1672 &26                 &10.5                 &1.0                 &3.8                &1.6               &1.0   &21       &8.2     &1\\
$\Omega(1^2P_{\frac{1}{2}^-})\pi^+$ &1957 &9.5                &3.8                &0.38                &2.0                &0.80              &0.50  &8.7      &3.6    &0.44\\
$\Omega(1^2P_{\frac{3}{2}^-})\pi^+$ &2012 &5.4                &2.2                &0.22                &1.2                &0.49              &0.31  &5.2      &2.1    &0.26\\
$\Omega(2^2S_{\frac{1}{2}^+})\pi^+$ &2232 &1.2                &5.0$\times10^{-1}$ &0.05                &3.9$\times10^{-1}$ &0.16              &0.01  &1.5      &6.3$\times10^{-1}$ &0.08\\
$\Omega(2^4S_{\frac{3}{2}^+})\pi^+$ &2159 &3.0                &1.2                &0.12                &0.8               &0.34              &0.21  &3.3      &1.4  &0.17\\
$\Omega(1^2D_{\frac{3}{2}^+})\pi^+$ &2245 &2.1$\times10^{-1}$ &8.4$\times10^{-2}$ &0.008               &6.7$\times10^{-2}$ &2.7$\times10^{-2}$&0.002 &2.6$\times10^{-1}$ &1.1$\times10^{-1}$ &0.01                     \\
$\Omega(1^2D_{\frac{5}{2}^+})\pi^+$ &2303 &1.3$\times10^{-2}$ &5.0$\times10^{-3}$ &5.0$\times10^{-4}$  &5.4$\times10^{-3}$ &2.0$\times10^{-3}$&1$\times10^{-3}$&1.9$\times10^{-2}$ &7.7$\times10^{-3}$ &9.4$\times10^{-4}$                          \\
$\Omega(1^4D_{\frac{1}{2}^+})\pi^+$ &2141 &3.3                &1.3                &0.13                &8.8$\times10^{-1}$ &0.36              &0.23&3.6        &1.5   &0.18 \\
$\Omega(1^4D_{\frac{3}{2}^+})\pi^+$ &2188 &2.3                &0.95                &0.09                &6.8$\times10^{-1}$ &0.28              &0.18&2.7        &1.1   &0.13       \\
$\Omega(1^4D_{\frac{5}{2}^+})\pi^+$ &2252 &3.3$\times10^{-3}$ &1.3$\times10^{-3}$ &1.3$\times10^{-4}$ &1.2$\times10^{-3}$ &4.5$\times10^{-4}$&2.8$\times10^{-4}$ &4.2$\times10^{-3}$ &1.7$\times10^{-3}$     &2.1$\times10^{-4}$                \\
$\Omega(1^4D_{\frac{7}{2}^+})\pi^+$ &2321 &3.2$\times10^{-3}$ &1.3$\times10^{-3}$ &1.3$\times10^{-4}$ &1.3$\times10^{-3}$ &5.1$\times10^{-4}$&3.2$\times10^{-4}$ &4.7$\times10^{-3}$ &1.9$\times10^{-3}$     &2.3$\times10^{-4}$                         \\
\hline\hline
\end{tabular}
\end{center}
\end{table*}

Considering $\Omega(2012)$ as the $\Omega (1^2P_{3/2^-})$ assignment, we have studied the
$\Omega_c^0\to \Omega^-(2012) \pi^+$ process, the results are listed in Table~\ref{result}.
It is found that the $\Omega_c$ baryon has a fairly large decay rate into $\Omega(2012)^- \pi^+$,
with the RPF or MHPF the branching fraction is predicted to be
\begin{eqnarray}\label{a3}
\mathcal{B}[\Omega_c^0\rightarrow \Omega(2012)^- \pi^+]&\simeq& 2.2
\times 10^{-3}.
\end{eqnarray}
Combining it with the branching fraction of $\mathcal{B}[\Omega_c^0\rightarrow \Omega^- \pi^+]$ obtained in Eq.(\ref{bb}),
we predict the relative ratio
\begin{eqnarray}\label{Hww}
R^{{\rm Th.}}_1=\frac{\mathcal{B}[\Omega_c^0\rightarrow
\Omega(2012)^- \pi^+]}{\mathcal{B}[\Omega_c^0\rightarrow \Omega^-
\pi^+]}\simeq 0.22,
\end{eqnarray}
which is in good agreement with experimental value $R^{\rm Exp.}_1 =
0.220\pm0.059(\mathrm{stat.})\pm0.035(\mathrm{syst.})$ that was
recently measured by the Belle Collaboration~\cite{Belle:2021gtf}.
According to the strong decay properties of $\Omega(2012)$ predicted
using the constituent quark model in
Refs.~\cite{Liu:2019wdr,Xiao:2018pwe}, branching fractions of
$\Omega(2012)$ decaying into $\Xi^0K^-$ and $\Xi^-\bar{K}^0$ are predicted
to be $\mathcal{B}[\Omega_c(2012)\to \Xi^0K^-]\simeq52\%$ and
$\mathcal{B}[\Omega_c(2012)\to \Xi^-\bar{K}^0 ]\simeq 48\%$, respectively.
Combining these strong branching fractions of $\Omega(2012)$ with
our predicted branching fractions for the weak decay processes
$\mathcal{B}[\Omega_c^0\to\Xi^0\bar{K}^-\pi^+/\Xi^-\bar{K}^0\pi^+/\Omega(2012)^-
\pi^+$] in Eqs.~(\ref{a1})-(\ref{a3}), one can obtain
\begin{eqnarray}\label{a4}
R^{\rm Th.}_2 &=& \frac{\mathcal{B}[\Omega_c^0\rightarrow \Omega(2012)\pi^+] \mathcal{B}[\Omega_c(2012)\rightarrow \Xi^0K^-]}{\mathcal{B}[\Omega_c^0\rightarrow \Xi^0\bar{K}^-\pi^+]} \nonumber \\
&& \simeq 0.09,\\
R^{\rm Th.}_3 &=& \frac{\mathcal{B}[\Omega_c^0\rightarrow
\Omega(2012)\pi^+] \mathcal{B}[\Omega_c(2012)\rightarrow
\Xi^-\bar{K}^0]}{\mathcal{B}[\Omega_c^0\rightarrow
\Xi^-\bar{K}^0\pi^+]} \nonumber \\
&& \simeq 0.05,
\end{eqnarray}
which are also consistent with the experimental values $R_2^{\rm
Exp.} = 0.096 \pm 0.032 (\mathrm{stat.}) \pm 0.018(\mathrm{syst.})$
and $R_3^{\rm Exp.} = 0.055 \pm 0.028(\mathrm{stat.}) \pm 0.007
(\mathrm{syst.})$ recently measured by the Belle
Collaboration~\cite{Belle:2021gtf}, respectively.
It should be mentioned that these predicted relative
ratios $R^{{\rm Th.}}_i$ ($i=1,2,3$) are nearly independent on the options of phase space
factor in the calculations.

Then we consider the weak decay rate of $\Omega_c$ into the other
$1P$-wave state $ \Omega (1^2P_{1/2^-}) $ by emitting a $\pi^+$
meson. The mass of $\Omega (1^2P_{1/2^-})$ is predicted to be $\sim
1950$ MeV within the Lattice QCD~\cite{Engel:2013ig} and the
relativized quark models~\cite{Capstick:1985xss,Faustov:2015eba}.
Experimentally, there seems to be a weak enhancement around 1950 MeV
in the $\Xi \bar{K}$ invariant mass distributions from the Belle
observations~\cite{Belle:2021gtf,Belle:2018mqs}, which may be a hint
of $\Omega (1^2P_{1/2^-})$. Hence, in the calculations the mass of
$\Omega (1^2P_{1/2^-})$ is taken to be 1957 MeV. If adopting the RPF or MHPF,
the branching fraction is predicted to be
\begin{eqnarray}\label{Hww}
\mathcal{B}[\Omega_c^0\rightarrow \Omega(1^2P_{1/2^-})
\pi^+]&\simeq& 3.8 \times 10^{-3},
\end{eqnarray}
which is about a factor of $5$ larger than that predicted with NRPF.
The predicted branching fraction $\mathcal{B}[\Omega_c^0\rightarrow \Omega(1^2P_{1/2^-})
\pi^+]$ should be slightly larger than that of the $\Omega(2012)^- \pi^+$
final states. The branching fraction ratio between
$\Omega_c^0\rightarrow \Omega(1^2P_{1/2^-}) \pi^+$ and
$\Omega_c^0\to \Omega^- \pi^+$ is predicted to be
\begin{eqnarray}\label{Hww}
\frac{\mathcal{B}[\Omega_c^0\to \Omega(1^2P_{1/2^-}) \pi^+]}{\mathcal{B}[\Omega_c^0\to \Omega^- \pi^+]}\simeq 0.38-0.50,
\end{eqnarray}
which is insensitive to options of the phase space
factor. Such a large relative branching ratio indicates that the other missing $1P$-wave state
$\Omega (1^2P_{1/2^-})$ has a good potential to be observed
in the weak decay process $\Omega_c^0\rightarrow \Omega(1^2P_{1/2^-}) \pi^+$.

According to the strong decay analysis in
Refs.~\cite{Liu:2019wdr,Xiao:2018pwe,Wang:2018hmi}, the decays of
$\Omega (1^2P_{1/2^-})$ should be nearly saturated by the $\Xi^0
K^-$ and $\Xi^- \bar{K}^0$ channels. Combined the strong decay
properties predicted within the chiral quark model in
Refs.~\cite{Liu:2019wdr,Xiao:2018pwe}, we can estimate the ratios
\begin{eqnarray}\label{a4}
\frac{\mathcal{B}[\Omega_c^0\to \Omega(1^2P_{1/2^-}) \pi^+] \mathcal{B}[\Omega(1^2P_{1/2^-})\to \Xi^0K^-]}{\mathcal{B}[\Omega_c^0\to \Xi^0K^-\pi^+]}\simeq16\%,\\
\frac{\mathcal{B}[\Omega_c^0\to\Omega(1^2P_{1/2^-}) \pi^+] \mathcal{B}[\Omega(1^2P_{1/2^-})\to\Xi^-\bar{K}^0]}{\mathcal{B}[\Omega_c^0\rightarrow \Xi^-\bar{K}^0\pi^+]}\simeq8\%,
\end{eqnarray}
which may provide useful references for future experiments.

To further explain the results of the  $ \Omega
(1^2P_{1/2^-}) $ and $\Omega (1^2P_{3/2^-}) $ states, we fit the $(\bar{K}\Xi)^-$ invariant mass spectrum of the
process $\Omega_c \rightarrow \pi^+\Omega^*(X)\rightarrow \pi^+ (\bar{K}\Xi)^-$
 measured by Belle Collaboration~\cite{Belle:2021gtf}. In our analysis, we adopt a relativistic Breit-Wigner function to describe the event distribution~\cite{ParticleDataGroup:2020ssz,Xie:2015lta,Xie:2018gbi,Zhang:2017eui}
\begin{equation}\label{dww}
\frac{dN}{dM_{(\bar{K}\Xi)^-}}= f_{BG}+C_R\sum_R\frac{M_{(\bar{K}\Xi)^-}^2\Gamma_{\pi^+\Omega^*(X)}(M_{(\bar{K}\Xi)^-})\Gamma_{(\Xi \bar{K})^-}(M_{(\bar{K}\Xi)^-})}{\left|M_{(\bar{K}\Xi)^-}^2-m_R^2+i m_R\Gamma_R\right|^2},
\end{equation}
where $M_{(\bar{K} \Xi)^-)}$ and $m_{R}$ stands for the invariant mass of $(\bar{K} \Xi)^-$ and the resonance mass of $\Omega^*(X)$, respectively.
$\Gamma_{\pi^+\Omega^*(X)}(M_{(\bar{K}\Xi)^-})$ and $\Gamma_{(\Xi \bar{K})^-}(M_{(\bar{K}\Xi)^-})$ are the partial decay widths of $\Omega_c^0\rightarrow \Omega^*(X) \pi^+$ and $\Omega^*(X)\rightarrow (\Xi K)^-$, respectively.
The total decay width $\Gamma_R$ are adopted as the predictions obtained in Ref.~\cite{Liu:2019wdr}, while $f_{BG}$ stands for the background contributions. In this work a linear background $f_{BG}=18.5$ (MeV$/c^2)^{-1}$  is adopted, which is determined by fitting the backgrounds taken in Ref.~\cite{Belle:2021gtf}. Finally $C_R$ is a global parameter related to the resonance production rates.

\begin{figure}[htbp]
\centering
\includegraphics[scale=0.40]{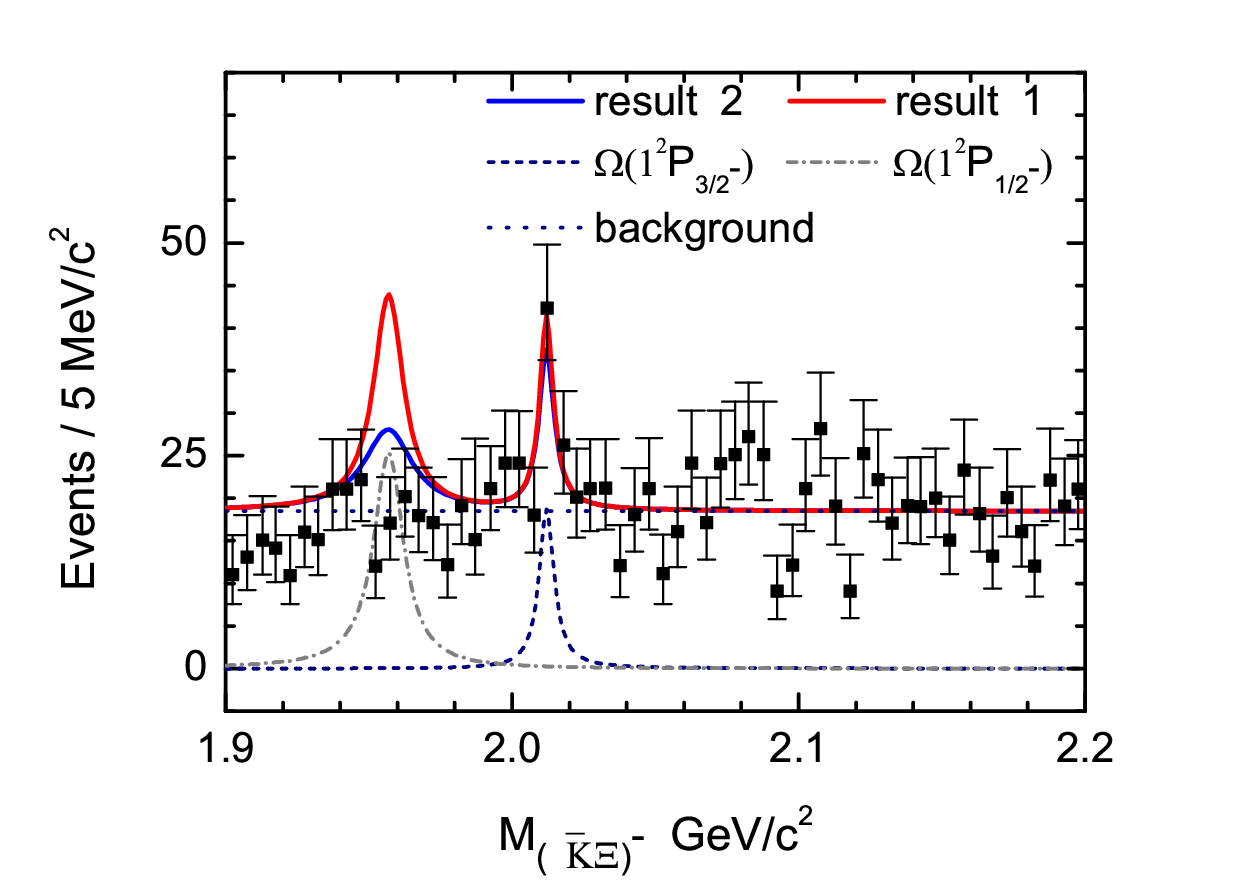}
\vspace{-0.5cm} \caption{The $(\bar{K}\Xi)^-$ invariant mass spectrum measured of the decay $\Omega_c \rightarrow \pi^+\Omega^*(X)\rightarrow \pi^+ ((\bar{K}\Xi)^-)$ by Belle Collaboration~\cite{Belle:2021gtf}
(solid squares) compared to the theoretical description with two
possible  $\Omega^-$(1P)-wave states, $ \Omega
(1^2P_{1/2^-}) $ and $\Omega (1^2P_{3/2^-}) $. Results 1  and 2 are fitting results of $ \Omega
(1^2P_{1/2^-}) $ with widths about 12.4 MeV and 20.0 MeV, respectively.}\label{KXinihe}
\end{figure}

In Fig.~\ref{KXinihe}, we shown our theoretical results for the $(\bar{K}\Xi)^-$ invariant mass distributions of the decay $\Omega_c \rightarrow \pi^+\Omega^*(X)\rightarrow \pi^+ ((\bar{K}\Xi)^-)$. The red curve has been adjusted to the strength of the experimental data of Belle Collaboration~\cite{Belle:2021gtf} at the peak around 2012 MeV by taking $C_R = 0.064$. Furthermore, the dashed curve stands for the resonance contribution of $\Omega(1^2P_{3/2^-})$ with $M_R = 2012$ MeV and $\Gamma_R =5.7$ MeV, while the dash-dotted curve stands for the $\Omega(1^2P_{1/2^-})$ contribution with $M_R = 1957$ MeV and $\Gamma_R =12.4$ MeV. From Fig.~\ref{KXinihe} one can easily find that the $\Omega(1^2P_{3/2^-})$ state has a significant contribution around 2012 MeV and the experimental data around that energy can be well reproduced. However, the contribution of the $\Omega(1^2P_{1/2^-})$ state is overestimated comparing with the experimental data around 1957 MeV. Yet, the quark model predicted widths for the $\Omega(1^2P_{1/2^-})$ and $\Omega(1^2P_{3/2^-})$ states have uncertainties, we perform a new calculation with a slightly large width $\Gamma_R = 20.0$ MeV for $\Omega(1^2P_{1/2^-})$ state, while we take the experimental value of $6.4$ MeV for $\Omega(1^2P_{3/2^-})$. The new theoretical results are also shown in Fig.~\ref{KXinihe} with blue curve, where we see that the signal of the $\Omega(1^2P_{1/2^-})$ is much suppressed. It is expected that more precise experimental data can be used to pin down the contribution of the $\Omega(1^2P_{1/2^-})$ state in future.

On the other hand, the $\Omega^0_c \to \pi^+\Omega(2012) \to \pi^+ K^-\Xi^0$ decay was investigated within the picture that the $\Omega(2012)$ is a molecular state in Ref.~\cite{Ikeno:2022jpe}, where the numerical results are also consistent with the experimental data. Indeed, we need further efforts to understand the nature of $\Omega(2012)$ state~\cite{Belle:2022mrg,Hu:2022pae}.

\subsection{$\Omega_c^0\rightarrow \Omega^-(1D) \pi^+$ }\label{1p1}

There are six $1D$-wave states, $ \Omega (1^2D_{3/2^+,5/2^+}) $ and
$\Omega (1^4D_{1/2^+,3/2^+,5/2^+,7/2^+}) $ according to the quark
quark model classification. Most of the predicted masses for the
$1D$-wave states lies in the mass range $\sim 2200 \pm 50$ MeV in
various quark models. Taking the mass recently predicted in
Ref.~\cite{Liu:2019wdr}, we calculate the weak decay properties for
the $\Omega_c^0\rightarrow \Omega^-(1D) \pi^+$ processes. Our
results are listed in Table~\ref{result}. It is seen that $\Omega_c$
has significant branching fractions decaying into the spin quartet
states $\Omega (1^4D_{1/2^+})$ and $\Omega (1^4D_{3/2^+})$. The
predicted branching fractions $\mathcal{B}[\Omega_c^0\rightarrow \Omega(1^4D_{1/2^+,3/2^+})\pi^+]$
can reach up to the order of $\sim O(10^{-4})-O(10^{-3})$.
With the RPF, their relative ratios to $\mathcal{B}[\Omega_c^0\rightarrow \Omega^- \pi^+]$
are predicted to be
\begin{eqnarray}\label{a9}
\frac{\mathcal{B}[\Omega_c^0\rightarrow \Omega(1^4D_{1/2^+}) \pi^+]}{\mathcal{B}[\Omega_c^0\rightarrow \Omega^- \pi^+]}\simeq 0.13,\\
\frac{\mathcal{B}[\Omega_c^0\rightarrow \Omega(1^4D_{3/2^+})
\pi^+]}{\mathcal{B}[\Omega_c^0\rightarrow \Omega^- \pi^+]}\simeq
0.09,
\end{eqnarray}
which are close to the results predicted with RPF and MHPF.
The predicted branching fractions and ratios are comparable with those of $\Omega_c$
decaying into the $\Omega(2012)\pi^+$ and $\Omega(1^2P_{1/2^-}) \pi^+$ channels. However, the decay rates of  $\Omega_c$ into the other four $1D$-wave states $ \Omega (1^2D_{3/2^+,5/2^+}) $ and $\Omega (1^4D_{5/2^+,7/2^+}) $ are  $\sim 1-3$
orders of magnitude smaller. The relatively large decay rates indicate that both $\Omega (1^4D_{1/2^+})$ and
$\Omega (1^4D_{3/2^+})$ has good potentials to be established by using the weak decay processes
$\Omega_c^0\to \Omega (1^4D_{1/2^+,3/2^+}) \pi^+$.

We further analyze the reasons of the small decay rates of
$\Omega_c^0\to \Omega (1^4D_{5/2^+,7/2^+}) \pi^+/\Omega (1^2D_{3/2^+,5/2^+})\pi^+$ compared with
that of $\Omega_c^0\to \Omega (1^4D_{1/2^+,3/2^+}) \pi^+$ as follows. We note that the helicity transition amplitudes
\begin{equation}\label{amplitude}
\mathcal{M}_{JJ_z;\frac{1}{2}-\frac{1}{2}}\propto \sum_{M_L+S_z=J_z} \langle L M_L S S_z|J J_z\rangle \langle \Psi_{NLM_L}^{\sigma} \chi_{S_z}^{\sigma} | \hat{O} | \Psi_{\Omega_c} \chi_{-\frac{1}{2}}^\lambda  \rangle,
\end{equation}
where $\Psi_{\Omega_c}$ ($\Psi_{NLM_L}$) and $\chi_{-\frac{1}{2}}^\lambda$ ($\chi_{S_z}^{\sigma}$) are the spacial and spin wave functions of the initial (final) baryons, respectively. For the decay processes involving the spin quartet
states $\Omega (1^4D_{1/2^+,3/2^+,5/2^+,7/2^+}) $, the decay amplitude is the sum of $c_1\langle \Psi_{221}^S \chi_{-3/2}^{S} | \hat{O} | \Psi_{\Omega_c} \chi_{-\frac{1}{2}}^\lambda  \rangle$ and $c_2\langle \Psi_{220}^S \chi_{-1/2}^{S} | \hat{O} | \Psi_{\Omega_c} \chi_{-\frac{1}{2}}^\lambda  \rangle$.
These two terms have strong constructive and destructive interference for the $\Omega_c^0\to \Omega (1^4D_{1/2^+,3/2^+}) \pi^+$
and $\Omega_c^0\to \Omega (1^4D_{5/2^+,7/2^+}) \pi^+$, respectively. Thus, the decay rates of
$\Omega_c^0\to \Omega (1^4D_{5/2^+,7/2^+}) \pi^+$ are strongly suppressed by
the destructive interference between the two terms of the helicity transition amplitude.
While for the decay processes involving the spin doublet $\Omega (1^2D_{3/2^+,5/2^+}) $,
the decay amplitudes are proportional to $\langle \Psi_{220}^{\rho,\lambda} \chi_{-1/2}^{\rho,\lambda} | \hat{O} | \Psi_{\Omega_c} \chi_{-\frac{1}{2}}^\lambda  \rangle$. In this term, the contribution from the part of the spin wave functions
is about a factor of $2-4$ smaller than that for the spin quartet
states. Thus, the decay rates of $\Omega_c^0\to \Omega (1^2D_{3/2^+,5/2^+}) \pi^+$ is
suppressed by the relative small overlapping of the spin wave functions of the initial
and final states.

According to the analysis of the strong decay
properties~\cite{Liu:2019wdr,Xiao:2018pwe}, the $\Omega
(1^4D_{1/2^+})$ state has a width of $\Gamma\simeq 42$ MeV, and
dominantly decays into the $\Xi \bar{K}$ channel with a branching
fraction $\sim 94\%$. While the $\Omega (1^4D_{3/2^+})$ has a width
of $\Gamma\simeq 31$ MeV, and dominantly decays into $\Xi \bar{K}$
with a branching fraction $\sim 64\%$. Thus, the $\Xi^0 K^-$ and
$\Xi^- \bar{K}^0$ final states can be used to look for the $\Omega
(1^4D_{1/2^+})$ and $\Omega (1^4D_{3/2^+})$ states if they are
produced by the $\Omega_c$ weak decays. For the $\Omega
(1^4D_{1/2^+})$ state, by combining the results of RPF we can estimate the following ratios
\begin{eqnarray}\label{a5}
\frac{\mathcal{B}[\Omega_c^0\to \Omega(1^4D_{1/2^+}) \pi^+] \mathcal{B}[\Omega(1^4D_{1/2^+})\to \Xi^0K^-]}{\mathcal{B}[\Omega_c^0\to \Xi^0\bar{K}^-\pi^+]}\simeq5\%,\\
\frac{\mathcal{B}[\Omega_c^0\to\Omega(1^4D_{1/2^+}) \pi^+] \mathcal{B}[\Omega(1^4D_{1/2^+})\to\Xi^-\bar{K}^0]}{\mathcal{B}[\Omega_c^0\rightarrow \Xi^-\bar{K}^0\pi^+]}\simeq3\%,
\end{eqnarray}
while for the $\Omega (1^4D_{3/2^+})$ state, we can estimate the following ratios
\begin{eqnarray}\label{a6}
\frac{\mathcal{B}[\Omega_c^0\to \Omega(1^4D_{3/2^+}) \pi^+] \mathcal{B}[\Omega(1^4D_{3/2^+})\to \Xi^0K^-]}{\mathcal{B}[\Omega_c^0\to \Xi^0\bar{K}^-\pi^+]}\simeq2\%,\\
\frac{\mathcal{B}[\Omega_c^0\to\Omega(1^4D_{3/2^+}) \pi^+] \mathcal{B}[\Omega(1^4D_{3/2^+})\to\Xi^-\bar{K}^0]}{\mathcal{B}[\Omega_c^0\rightarrow \Xi^-\bar{K}^0\pi^+]}\simeq1\%.
\end{eqnarray}
The above predicted ratios are less dependent on the options of the phase space factor.

\begin{figure}[htbp]
\centering
\includegraphics[scale=0.42]{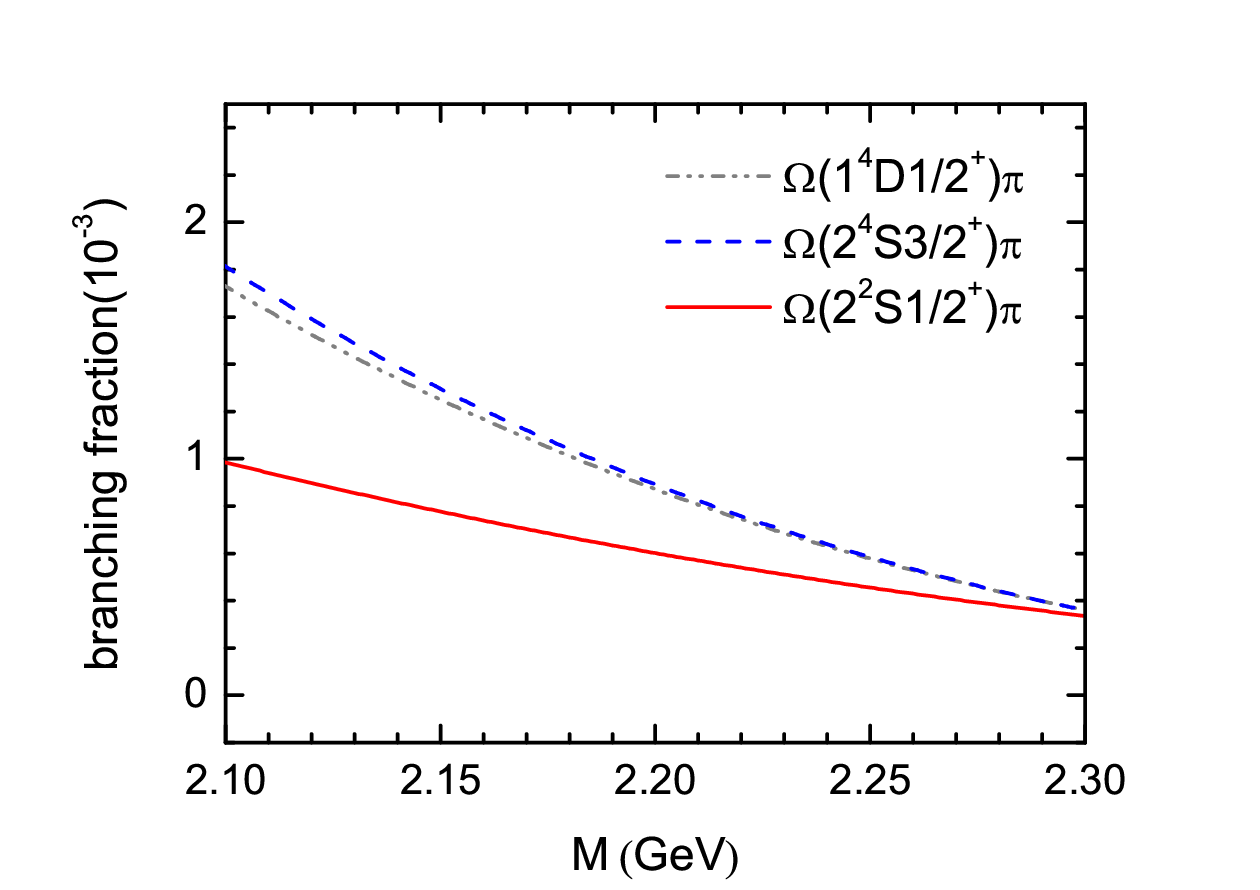}
\vspace{-0.5cm} \caption{The branching fraction of the
$\Omega_c^0\rightarrow \Omega(1^4D_{1/2^+},1^4D_{3/2^+},2^2S_{1/2^+},2^4S_{3/2^+}) \pi^+$ as a function of mass
of the final state $\Omega^*(X)$. It should be noted that since the results of $1^4D_{1/2^+}$ and $1^4D_{3/2^+}$ are the same, we omit
the results of $1^4D_{3/2^+}$ here.}\label{1Dwave}
\end{figure}

Finally, it should be pointed out that the predicted masses of the
$1D$-wave $\Omega$ states have some model dependencies. To see the
effects from the mass uncertainties of the $1D$-wave $\Omega$ states
on our predicted weak decay properties, we plot the weak branching fractions of $\Omega^0_c \to \pi^+ \Omega^*(X)$ as
functions of the masses of the $1D$-wave $\Omega$ excited state in their possible range $M \in (2.1-2.3)$ GeV in Fig.~\ref{1Dwave}. It is seen that in the most possible mass range $\sim 2200 \pm 50$ MeV, the upper limit of our predicted partial widths is about a factor of $2$ larger than that of the lower limit.

\subsection{$\Omega_c^0\rightarrow \Omega(2S) \pi^+$}

In the constituent quark model, there are two $2S$-wave states
$\Omega(2^2S_{1/2^+})$ and $\Omega(2^4S_{3/2^+})$. There are large
uncertainties in the predictions of their masses in various quark
models. The predicted masses scatter in the range of $\sim2.10 \sim
2.30$ GeV. In Fig.~\ref{1Dwave}, by using the RPF we plot the weak decay widths of
the $\Omega_c^0\rightarrow \Omega(2^2S_{1/2^+}/2^4S_{3/2^+}) \pi^+$
processes as functions of the masses of the $2S$-wave $\Omega$
states. It is seen that in the mass range $2100 \sim 2300$ MeV, for
the weak decay process $\Omega_c^0\rightarrow \Omega(2^2S_{1/2^+})
\pi^+$, the partial decay width is predicted to be
$\Gamma[\Omega_c^0\rightarrow \Omega(2^2S_{1/2^+}) \pi^+]\simeq
(1.2\pm 0.45)\times10^{-15}$ GeV, the branching fraction can reach
up to
\begin{eqnarray}\label{a10}
\mathcal{B}[\Omega_c^0\to \Omega(2^2S_{1/2^+}) \pi^+]&\simeq& (0.50
\pm 0.18)\times 10^{-3}.
\end{eqnarray}
Combined with the predicted branching fraction
$\mathcal{B}[\Omega_c^0\rightarrow \Omega^- \pi^+]\simeq 10\%$, we
obtain the relative branching ratio
\begin{eqnarray}\label{Hww}
\frac{\mathcal{B}[\Omega_c^0\rightarrow \Omega(2^2S_{1/2^+}) \pi^+]}{\mathcal{B}[\Omega_c^0\rightarrow \Omega^- \pi^+]}\simeq 0.05\pm0.02
\end{eqnarray}
The production rate of $\Omega(2^2S_{1/2^+})$ via the $\Omega_c$
weak decay is about a factor of $5-6$ smaller than that of
$\Omega(2012)$. Due to the large decay rate into the $\Xi(1530)
\bar{K}$ channel~\cite{Liu:2019wdr,Xiao:2018pwe}, the
$\Omega(2^2S_{1/2^+})$ state is suggested to be searched in the
decay chain $\Omega_c^0\rightarrow \Omega(2^2S_{1/2^+}) \pi^+ \to
(\Xi(1530) K)^-\pi^+\to (\Xi\pi K)^-\pi^+$ in future experiments.

For the other weak decay process $\Omega_c^0 \to \Omega(2^4S_{3/2^+}) \pi^+$,
by using the RPF the partial decay width is predicted to be
$\Gamma[\Omega_c^0\rightarrow \Omega(2^4S_{3/2^+}) \pi^+]\simeq (3.0\pm 1.6)\times10^{-15}$ GeV,
the branching fraction can reach up to
\begin{eqnarray}\label{a10}
\mathcal{B}[\Omega_c^0\to \Omega(2^4S_{3/2^+}) \pi^+]&\simeq& (1.2
\pm 0.6)\times 10^{-3}.
\end{eqnarray}
Similarly, the relative branching ratio is predicted to be
\begin{eqnarray}\label{Hww}
\frac{\mathcal{B}[\Omega_c^0\rightarrow \Omega(2^4S_{3/2^+}) \pi^+]}{\mathcal{B}[\Omega_c^0\rightarrow \Omega^- \pi^+]}\simeq 0.12\pm 0.06.
\end{eqnarray}
The production rate of $\Omega(2^4S_{3/2^+})$ via the $\Omega_c$
weak decay is comparable with that of $\Omega(2^2S_{1/2^+})$. The
dominant decay mode of $\Omega(2^4S_{3/2^+})$ is the $\Xi(1530)
\bar{K}$ channel, and one can look for it in the decay chain
$\Omega_c^0\to \Omega(2^4S_{3/2^+}) \pi^+ \to (\Xi(1530)
\bar{K})^-\pi^+\to (\Xi\pi \bar{K})^-\pi^+$.

\section{Summary}\label{SUM}

In this work, we calculate the Cabbibo-favored weak decay processes
$\Omega_c \rightarrow \Omega^{(*)}(X)\pi^+$ within a constituent
quark model. Our predicted branching fraction
$\mathcal{B}[\Omega_c^0\rightarrow \Omega^- \pi^+]\simeq 1.0\%$
which is in agreement with the early predictions in orders in Refs.~\cite{Cheng:1996cs,Korner:1992wi}.  Considering the newly observed
$\Omega(2012)$ resonance as the conventional $\Omega (1^2P_{3/2^-})$
state, it is found that the measured ratio $\mathcal{B}[\Omega_c\to
\Omega(2012)\pi^+ \to (\Xi\bar{K})^-\pi^+ ]/\mathcal{B}[\Omega_c\to
\Omega \pi^+]=0.220\pm0.059(\mathrm{stat.})\pm0.035(\mathrm{syst.})$
at Belle can be well understood within our model calculations here.
The production potentials of the missing low-lying $1P$-, $2S$-, and
$1D$-wave resonances $\Omega^*(X)$ via the hadronic weak decays of
$\Omega_c$ are discussed as well. Our main conclusions are
summarized as follows.

i) The missing $1P$-wave state $\Omega (1^2P_{1/2^-})$ has a large
potential to be observed in the decay chain $\Omega_c^0\to\Omega
(1^2P_{1/2^-})\pi^+ \to (\Xi \bar{K})^-\pi^+$. The production rate
of $\Omega(1^2P_{1/2^-})$ via the hadronic weak decays of $\Omega_c$
is even slightly larger than that of $\Omega(2012)$.

ii) For the $1D$-wave $\Omega$ states, we find that both $\Omega
(1^4D_{1/2^+})$ and $\Omega (1^4D_{3/2^+})$ have fairly large
production rates via the $\Omega_c \to \Omega (1^4D_{1/2^+})\pi^+$
and $\Omega_c \to \Omega (1^4D_{3/2^+})\pi^+$ processes,
respectively. Their production rates via the hadronic weak decays of
$\Omega_c$ are comparable with those of the $1P$-wave $\Omega$
states. Both $\Omega (1^4D_{1/2^+})$ and $\Omega (1^4D_{3/2^+})$ are
most likely to be observed in the process $\Omega_c^0\to \Omega
(1^4D_{1/2^+,3/2^+}) \pi^+\to (\Xi \bar{K})^- \pi^+$.

iii) The $2S$ states $\Omega(2^2S_{1/2^+})$ and
$\Omega(2^4S_{3/2^+})$ also have fairly large production rates via
the hadronic weak decays of $\Omega_c$. Their production rates are
about a factor of $5-6$ smaller than that of $\Omega(2012)$. Both
$\Omega(2^2S_{1/2^+})$ and $\Omega(2^4S_{3/2^+})$ dominantly decay
into the $\Xi(1530)K$ channel, thus, they can be looked for in the
decay chains $\Omega_c^0\to \Omega(2^2S_{1/2^+}) \pi^+/
\Omega(2^4S_{3/2^+}) \pi^+ \to (\Xi(1530) \bar{K})^-\pi^+\to (\Xi\pi
\bar{K})^-\pi^+$.

Finally, it should be mentioned that our predicted
partial widths for the weak decay processes $\Omega_c \to
\Omega^{(*)}\pi^+$ may have a large uncertainties due to relativistic effects.
To roughly see the uncertainties from the relativistic corrections,
we perform our calculations with the three typical phase space options,
the relativistic phase space, the nonrelativistic phase space and the
``mock-hadron'' phase space. The predicted partial widths with
the nonrelativistic phase space are a factor of
$\sim2-6$ smaller those calculated with the usual relativistic phase space
and the ``mock-hadron'' phase space.

\section*{Acknowledgements }

This work is supported by the National Natural Science Foundation of
China (Grants Nos.12205026, 12175065, 12075288, U1832173, 11775078, 11705056,
11735003, and 11961141012), and  Applied Basic Research Program of Shanxi Province, China under Grant No.202103021223376. Qi-Fang L\"{u} is partly supported by
the State Scholarship Fund of China
Scholarship Council under Grant No. 202006725011. Ju-Jun Xie is also
supported by the Youth Innovation Promotion Association CAS.

%\end{spacing}

\end{document}